\documentclass[10pt,letterpaper]{article}

\usepackage{cogsci}
\usepackage{pslatex}
\usepackage{apacite}
\usepackage{graphicx}

\usepackage[justification=centering]{caption}
\usepackage{array}
\usepackage{ragged2e}
\usepackage{epstopdf}

\usepackage{amsmath}%
\usepackage{amsfonts}%
\usepackage{amssymb}%

\title{Computational Evidence that Self-regulation of Creativity is Good for Society}
 
\author{{\large \bf Liane Gabora (liane.gabora@ubc.ca)}\\
  University of British Columbia\\
  Department of Psychology, Okanagan campus, Arts Building, 3333 University Way\\
  Kelowna BC, V1V 1V7, CANADA\\
 \AND {\large \bf Simon Tseng (s.tseng@alumni.ubc.ca)}\\
  University of British Columbia \\
  Department of Engineering, 5000-2332 Main Mall\\
  Vancouver BC,V6T 1Z4, CANADA\\
}

\begin{document}

\maketitle

\begin{abstract}
Excess individual creativity can be detrimental to society because creators invest in unproven ideas at the expense of propagating proven ones. Moreover, a proportion of individuals can benefit from creativity without being creative themselves by copying creators. We hypothesized that (1) societies increase their rate of cultural evolution by tempering the novelty-generating effects of creativity with the novelty-preserving effects of imitation, and (2) this is carried out by selectively rewarding and punishing creativity according to the value of the individuals' creative outputs. We tested this using an agent-based model of cultural evolution in which each agent self-regulated its invention-to-imitation ratio as a function of the fitness of its cultural outputs. In self-regulating societies, agents segregated into creators and imitators. The mean fitness of cultural outputs was higher than in non-self-regulating societies, and changes in diversity were rapider and more pronounced. We discuss limitations and possible social implications of our findings.
\\

\end{abstract}

\section{Introduction}

It is commonly assumed that creativity is desirable, and the more creative one is, the better. Our capacity for self-expression, problem solving, and making aesthetically pleasing artifacts, all stem from our creative abilities. However, individuals often claim that their creativity is stifled by social norms, policies, and institutions. Moreover, our educational systems do not appear to prioritize the cultivation of creativity, and in some ways discourage it. 

Perhaps there is an adaptive value to these seemingly mixed messages that society sends about the social desirability of creativity. Perhaps what is best for society is that individuals vary widely with respect to how creative they are, so as to ensure that the society as a whole both generates novel variants, and preserves the best of them. This paper provides a computational test of the hypothesis that society as a whole benefits when individuals can vary how creative they are in response to the perceived effectiveness of their ideas.

\subsection{Definition and Key Features of Creativity}

There are a plethora of definitions of creativity in the literature; nevertheless, it is commonly accepted that a core characteristic of creativity is the production of an idea or product that meets two criteria: originality or \emph{novelty}, and appropriateness, adaptiveness, or \emph{usefulness, i.e.,} relevance to the task at hand \cite{Guilford1950,Moran2011}. Not only are humans individually creative, but we build on each other's ideas such that over centuries, art, science, and technology, as well as customs and folk knowledge, can be said to evolve. This cumulative building of new innovations on existing products is sometimes referred to as the ratchet effect \cite{Tomasello1993}.  Creativity has long been associated with personal fulfillment \cite{May1975,Rogers1959}, self-actualization \cite{Maslow1959}, and maintaining a competitive edge in the marketplace. Thus it is often assumed that more creativity is necessarily better. 

However, there are significant drawbacks to creativity \cite{CropleyCropleyKaufmanRunco2010,Ludwig1995}. Generating creative ideas is difficult and time consuming, and a creative solution to one problem often generates other problems, or has unexpected negative side effects that may only become apparent after much effort has been invested. Creativity is correlated with rule bending, law breaking, and social unrest \cite{SternbergLubart1995,Sulloway1996}, aggression \cite{TacherReaddick2006}, group conflict \cite{TroyerYoungreen2009}, and dishonesty \cite{GinoAriely2012}. Creative individuals are more likely to be viewed as aloof, arrogant, competitive, hostile, independent, introverted, lacking in warmth, nonconformist, norm doubting, unconscientious, unfriendly \cite{BateyFurnham2006,QianPluckerShen2010,TreffingerYoungShelbyShepardson2002}. They tend to be more emotionally unstable, and more prone to affective disorders such as depression and bipolar disorder, and have a higher incidence of schizophrenic tendencies, than other segments of the population \cite{Andreason1987,Eysenck1993,Flaherty2005}. They are also more prone to drug and alcohol abuse, as well as suicide \cite{Jamieson1993,Goodwin1998,Rothenberg1990,Kaufman2003}. This suggests that there is a cost to creativity, both to the individual and to society. 

\subsection{Balancing Novelty with Continuity}

Given the correlation between creativity and personality traits that are potentially socially disruptive, it is perhaps fortunate that in a group of interacting individuals, not all of them need be particularly creative for the benefits of creativity to be felt throughout the group. The rest can reap the rewards of the creator's ideas by copying them, buying from them, or simply admiring them. Few of us know how to build a computer, or write a symphony, but they are nonetheless ours to use and enjoy. Of course, if everyone relied on the strategy of imitating others rather than coming up with their own ideas, the generation of cultural novelty would grind to a halt. On the other hand, if everyone were as creative as the most creative amongst us, the frequency of the above-mentioned antisocial tendencies of creative people might be sufficiently high to interfere with cultural stability; \emph{i.e.,} the perpetuation of cultural continuity. It is well known in theoretical biology that both novelty and continuity are essential for evolution, that is, for cumulative, open-ended, adaptive change over time. 

This need for both novelty and continuity was demonstrated in an agent-based model of cultural evolution \cite{Gabora1995}. Novelty was injected into the artificial society through the invention of new actions, and continuity was preserved through the imitation of existing actions. When agents never invented, there was nothing to imitate, and there was no cultural evolution at all. If the ratio of invention to imitation was even marginally greater than 0, not only was cumulative cultural evolution possible, but eventually all agents converged on optimal cultural outputs. When all agents always invented and never imitated, the mean fitness of cultural outputs was also sub-optimal because fit ideas were not dispersing through society. The society as a whole performed optimally when the ratio of creating to imitating was approximately 2:1. Although results obtained with a simple computer model may have little bearing on complex human societies, the finding that extremely high levels of creativity can be detrimental to the society suggests that there may be an adaptive value to society's ambivalent attitude toward creativity. 
% Without invention there is nothing to imitate, but imitation helps preserve fit inventions. 
% Moreover, creative individuals across a range of domains report that they intersperse prolonged immersion in a task with consultation with an expert or discussion with friends or colleagues, thus this result appears to correspond with anecdotal reports of creative individuals. 

This suggested that society as a whole might benefit from a distinction between the conventional workforce and what has been called a ``creative class" \cite{Florida2002} 
This was investigated in the model by introducing two types of agents: imitators, that only obtained new actions by imitating neighbors, and creators, that obtained new actions either by inventing or imitating \cite{GaboraFirouzi2012}. It was possible to vary the probability that creators create versus imitate; thus, whereas a given agent was either a creator or an imitator throughout the entire run, the proportion of creators innovating or imitating in a given iteration fluctuated stochastically. The mean fitness of ideas across the artificial society was highest when not all agents were creators. Specifically, there was a tradeoff between $C$, the proportion of creators to imitators in the society, and $p$, how creative the creators were). This provided further support for the hypothesis that society as a whole functions optimally when creativity is tempered with continuity. 

We then hypothesized that society as a whole might perform even better if individuals are able to adjust how creative they are over time in accordance with their perceived creative success. This could occur through mechanisms such as selective ostracization of deviant behaviour unless accompanied by the generation of valuable cultural novelty, and encouraging of successful creators. In this way society might self-organize into a balanced mix of novelty generating creators and continuity perpetuating imitators, both of which are necessary for cumulative cultural evolution. 
A first step in investigating this hypothesis was to determine whether it is algorithmically possible to increase the mean fitness of ideas in a society by enabling them to self-regulate how creative they are.

\section{The Computational Model}

We investigated this using an agent-based model of cultural evolution referred to as ``EVOlution of Culture'', abbreviated EVOC. 
% To our knowledge, EVOC is the only computational model that enables one to create an artificial world with agents of varying levels of creativity and observe the effect of varying creativity on mean fitness and diversity of ideas in the artificial society. 
It uses neural network based agents that (1) invent new ideas, (2) imitate actions implemented by neighbors, (3) evaluate ideas, and (4) implement successful ideas as actions. EVOC is an elaboration of Meme and Variations, or MAV \cite{Gabora1995}, the earliest computer program to model culture as an evolutionary process in its own right, as opposed to modeling the interplay of cultural and biological evolution\footnote{The approach can thus be contrasted with computer models of how individual learning affects biological evolution \cite{Best1999,Higgs2000,HintonNowlan1987,HutchinsHazelhurst1991}.}. 
% The approach was inspired by the genetic algorithm \cite{Holland1975}, or GA. The GA is a search technique that finds solutions to complex problems by generating a population of candidate solutions through processes akin to mutation and recombination, selecting the best, and repeating until a satisfactory solution is found. 
The goal behind MAV, and also behind EVOC, was to distil the underlying logic of not biological evolution but cultural evolution, \emph{i.e.,} the process by which ideas adapt and build on one another in the minds of interacting individuals. Agents do not evolve in a biological sense, for they neither die nor have offspring, but do in a cultural sense, by generating and sharing ideas for actions. In cultural evolution, the generation of novelty takes place through invention
%instead of through mutation and recombination as in biological evolution
, and the differential replication of novelty takes place through imitation
%instead of through reproduction with inheritance as in biological evolution
. EVOC was originally developed to investigate the similarities and differences between biological and cultural evolution, and it has been used to address such questions as how does the presence of leaders or barriers to the diffusion of ideas affect cultural evolution.

We now summarize briefly the architecture of EVOC in sufficient detail to explain our results; for further details we refer the reader to previous publications \cite{Gabora1995,GaboraChiaFirouzi2013,GaboraFirouzi2012}.

\subsection{Agents}

Agents consist of (1) a neural network, which encodes ideas for actions and detects trends in what constitutes a fit action, (2) a `perceptual system', which observes and evaluates neighbours' actions, and (3) a body, consisting of six body parts which implement actions. The neural network is composed of six input nodes and six corresponding output nodes that represent concepts of body parts (LEFT ARM, RIGHT ARM, LEFT LEG, RIGHT LEG, HEAD, and HIPS), and seven hidden nodes that represent more abstract concepts (LEFT, RIGHT, ARM, LEG, SYMMETRY, OPPOSITE, and MOVEMENT). Input nodes and output nodes are connected to hidden nodes of which they are instances (\emph{e.g.,} RIGHT ARM is connected to RIGHT.) Each body part can occupy one of three possible positions: a neutral or default position, and two other positions, which are referred to as active positions. Activation of any input node activates the MOVEMENT hidden node. Same-direction activation of symmetrical input nodes (\emph{e.g.,} positive activation -- which represents upward motion -- of both arms) activates the SYMMETRY node. 
%In the experiments reported here the OPPOSITE hidden node was not used.

\subsection{Invention}

An idea for a new action is a pattern consisting of six elements that dictate the placement of the six body parts. Agents generate new actions by modifying their initial action or an action that has been invented previously or acquired through imitation. During invention, the pattern of activation on the output nodes is fed back to the input nodes, and invention is biased according to the activations of the SYMMETRY and MOVEMENT hidden nodes. (Were this not the case there would be no benefit to using a neural network.) To invent a new idea, for each node of the idea currently represented on the input layer of the neural network, the agent makes a probabilistic decision as to whether the position of that body part will change, and if it does, the direction of change is stochastically biased according to the learning rate. If the new idea has a higher fitness than the currently implemented idea, the agent learns and implements the action specified by that idea. When ``chaining'' is turned on, an agent can keep adding new sub-actions and thereby execute a multi-step action, so long as the most recently-added sub-action is both an optimal sub-action and different from the previous sub-action of that action \cite{GaboraChiaFirouzi2013}.

\subsection{Imitation}

The process of finding a neighbour to imitate works through a form of lazy (non-greedy) search. The imitating agent randomly scans its neighbours, and adopts the first action that is fitter than the action it is currently implementing. If it does not find a neighbour that is executing a fitter action than its own current action, it continues to execute the current action. 

%added 2nd sentence -LC
\subsection{Evaluation: The Fitness Function}

Following \citeA{Holland1975}, we refer to the success of an action in the artificial world as its \emph{fitness}, with the caveat that unlike its usage in biology, here the term is unrelated to number of offspring (or ideas derived from a given idea). The fitness function used in these experiments rewards activity of all body parts except for the head, and symmetrical limb movement. Fitness of a single-step action $F_n$ is determined as per {Eq.}~\ref{eq:fitnesssingle}. Total body movement, $m$, is calculated by adding the number of active body parts, \emph{i.e.,} body parts not in the neutral position. 

\begin{equation}
F_{n} = m + 1.5 (s_a+s_t)+2(1-m_h)
\label{eq:fitnesssingle}
\end{equation}
$s_a = 1$ if arms move symmetrically; 0 otherwise \\
$s_t = 1$ if legs move symmetrically; 0 otherwise \\
$m_h = 1$ if head is stationary; 0 otherwise \\

Note that there are multiple optima. (For example an action can be optimal if either both arms move up or if both arms move down.) The fitness $F_c$ of a multi-step action with $n$ chained single-step actions (each with fitness $F_n$) is calculated by {Eq.}~\ref{eq:fitnesschained}.

\begin{equation}
F_c= \sum\limits_{k = 1}^n {\dfrac{F_n}{1.2^{n-1}}} 
\label{eq:fitnesschained}
\end{equation}

\subsection{Learning}

Invention makes use of the ability to detect, learn, and respond adaptively to trends. Since no action acquired through imitation or invention is implemented unless it is fitter than the current action, new actions provide valuable information about what constitutes an effective idea. Knowledge acquired through the evaluation of actions is translated into educated guesses about what constitutes a successful action by updating the learning rate. For example, an agent may learn that more overall movement tends to be either beneficial (as with the fitness function used here) or detrimental, or that symmetrical movement tends to be either beneficial (as with the fitness function used here) or detrimental, and bias the generation of new actions accordingly.

\subsection{The Artificial World}

These experiments used a default artificial world: a toroidal lattice with 1024 cells each occupied by a single, stationary agent, and a von Neumann neighborhood structure. Creators and imitators were randomly dispersed. 

\subsection{A Typical Run}

Fitness and diversity of actions are initially low because all agents are initially immobile, implementing the same action, with all body parts in the neutral position. Soon some agent invents an action that has a higher fitness than immobility, and this action gets imitated, so fitness increases. Fitness increases further as other ideas get invented, assessed, implemented as actions, and spread through imitation. The diversity of actions increases as agents explore the space of possible actions, and then decreases as agents hone in on the fittest actions. Thus, over successive rounds of invention and imitation, the agents' actions improve. EVOC thereby models how ``descent with modification'' occurs in a purely cultural context. 

\section{Method}
 
To test the hypothesis that the mean fitness of cultural outputs across society increases faster with social regulation (SR) than without it, we increased the relative frequency of invention for agents that generated superior ideas, and decreased it for agents that generated inferior ideas. To implement this the computer code was modified as follows. Each iteration, for each agent, the fitness of its current action relative to the mean fitness of actions for all agents at the previous iteration was assessed. Thus we obtained the relative fitness ($RF$) of its cultural output. An agent's personal probability of creating, $p(C)$, was a function of $RF$, determined as follows:

\begin{equation}
	p(C)_{n} = p(C)_{n-1} \times RF_{n-1}
\label{eq:socialregulation}
\end{equation}
 
The probability of imitating, $p(I)$, was 1 - $p(C)$. Thus when SR was on, if relative fitness was high the agent invented more, and if it was low the agent imitated more. $p(C)$ was initialized at 0.5 for both SR and non-SR societies. We compared runs with SR to runs without it. 

\section{Results}

All data are averages across 250 runs. The mean fitness of the cultural outputs of societies with SR (the ability to self-regulate inventiveness as a function of inventive success) was higher than that of societies without SR, as shown in Figure 1. The fact that ideas got increasingly fitter over the course of a run was due to the fact that chaining was turned on; a fit action could always be made fitter by adding another sub-action. When chaining was turned off, societies with SR also outperformed those without it, except that in both SR an non-SR societies the mean fitness plateaued when all agents converged on optimally fit ideas (not shown).

\begin{figure}[ht]
\centering
\includegraphics[width=0.95\columnwidth]{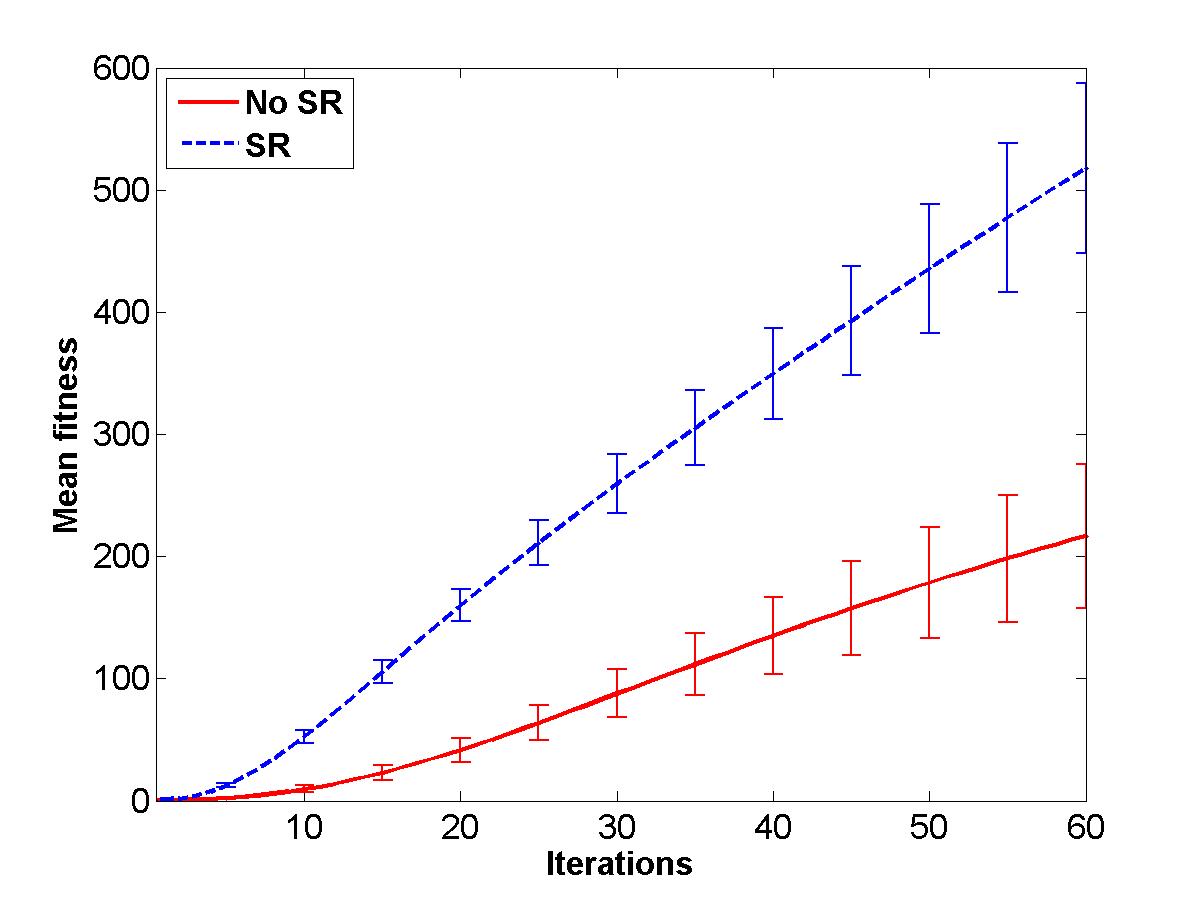}
\caption{This graph plots the mean fitness of implemented actions across all agents over the duration of the run with and without social regulation.}
\label{fig:F2-fitness-self-mod-soc}
\vskip -0.1in
\end{figure}

The typical pattern was observed with respect to the diversity, or number of different ideas: an increase as the space of possibilities is explored followed by a decrease as agents converge on fit actions. However, this pattern occurred earlier, and was more pronounced, in societies with SR than in societies without it, as shown in Figure 2. Inferior inventors tended to come up with the same ideas and thus discouraging them did not lower diversity, while the ideas of superior inventors tended to go off in different directions, thereby increasing diversity. Note that, with chaining turned on, although the number of different actions decreases, the agents do not converge on a static set of actions; the set of implemented actions changes continuously as they find new, fitter actions. 

\begin{figure}[ht]
\centering
\includegraphics[width=0.95\columnwidth]{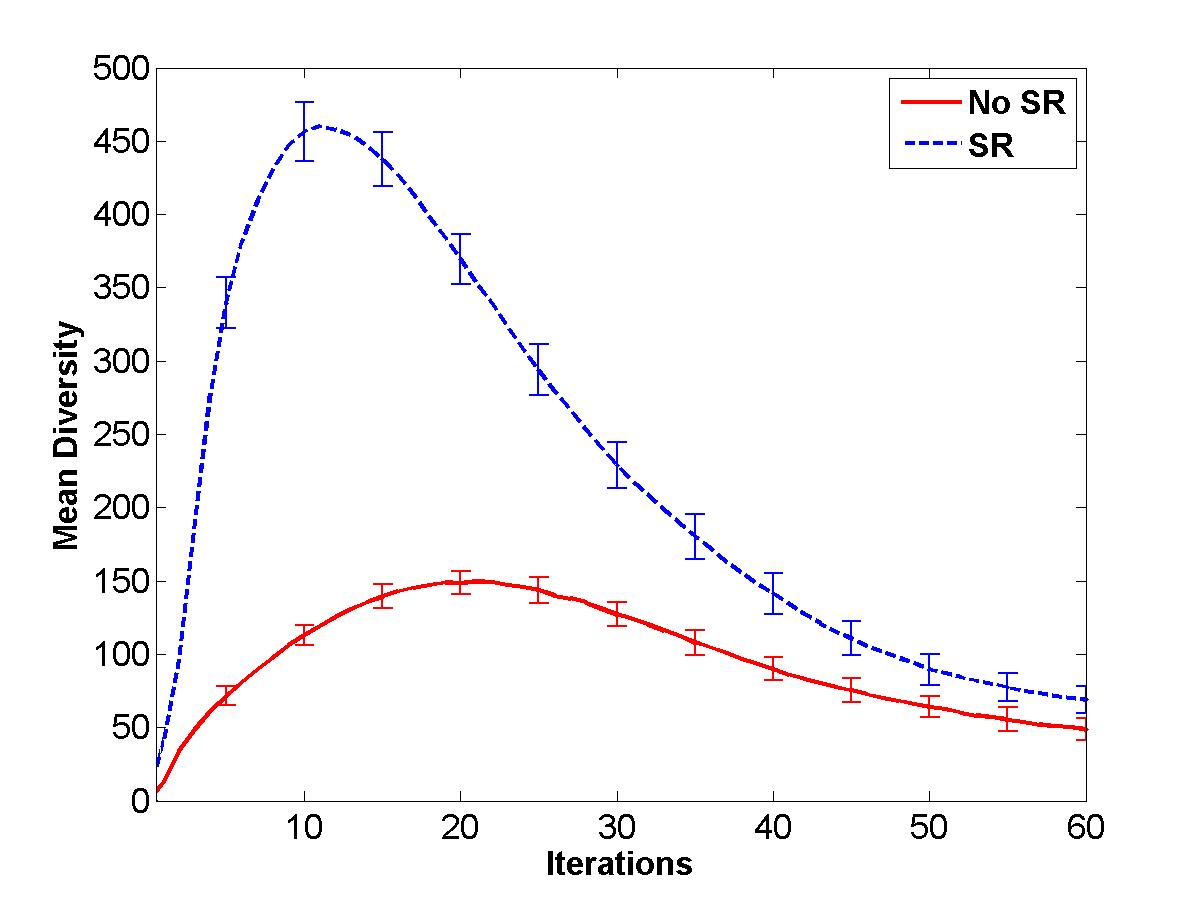}
\caption{This graph plots the diversity, or total number of different actions, across all agents over the duration of the run with and without social regulation.}
\label{fig:F3-diversity-self-mod-soc}
\vskip -0.1in
\end{figure}

Societies with SR ended up separating into two distinct groups: one that primarily invented, and one that primarily imitated, as illustrated in Figure 3. Thus the observed increase in fitness can indeed be attributed to increasingly pronounced  individual differences in their degree of creative expression over the course of a run. Agents that generated superior cultural outputs had more opportunity to do so, while agents that generated inferior cultural outputs became more likely to propagate proven effective ideas rather than reinvent the wheel. 

\begin{figure}[!htb]
\centering
\includegraphics[width=0.82\columnwidth]{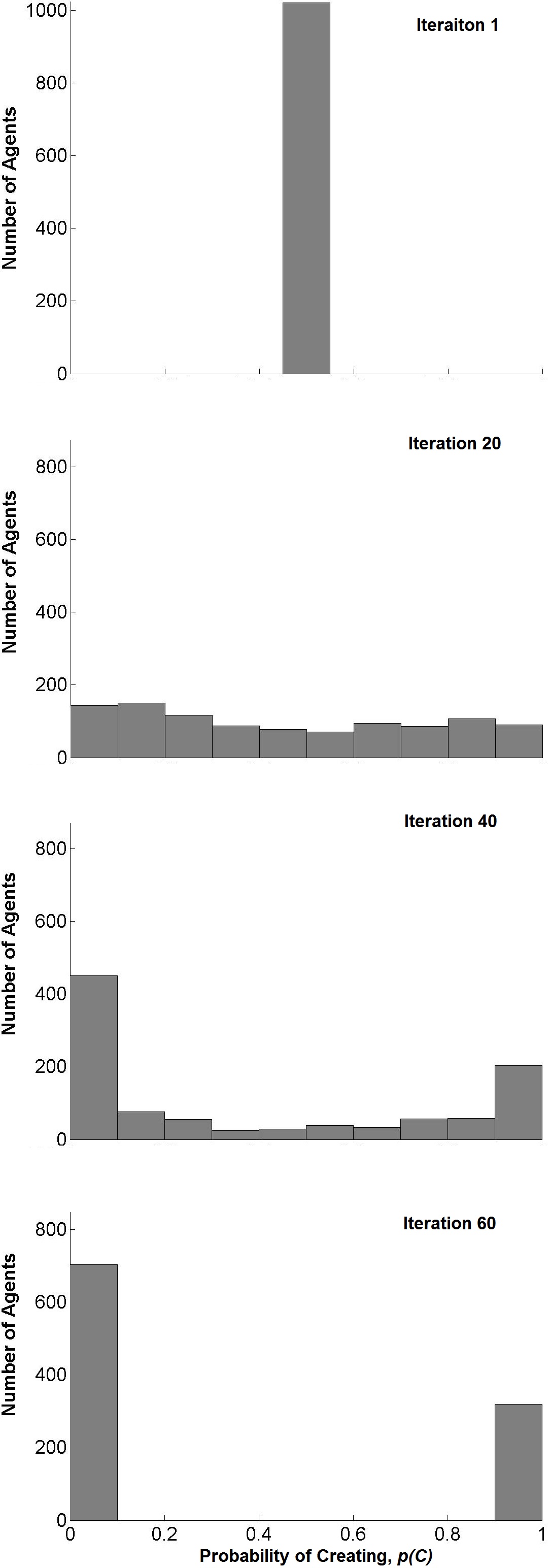}
\caption{At the beginning of the run (top) all agents created and imitated with equal probability. Midway through the run their $p(C)$ values were distributed along the range of values from 0 to 1. By the end of the run (bottom) they had segregated into imitators (with $p(C)$ from 0 to 0.1) and creators (with $p(C)$ from 0.9 to 1).}
%Figure1b(middle) shows a good majority of the agents have already taken to the extreme ends of the invention-to-imitation ratio while a portion of the agents are still in between at iteration 20. Figure1c(bottom) shows All the agents are now either spending the majority of iterations inventing or spending the majority of iterations imitating at iteration 60.}
\label{fig:F1-agent-distribution}
\vskip -0.1in
\end{figure}

\section{Discussion}
%
%This investigation yielded results that contradict the widespread assumption that creativity is necessarily desirable. The model is highly idealized, and caution must be taken in extrapolating to human societies. The PIV results assume that creators avoid input from neighbors if doing so would maximize the fitness of their actions. In reality, creative individuals may not behave so rationally. However, the PIV results were corroborated by the TTT results, indicating that the basic pattern does not depend on the assumption of economic rationality. 
%

Should society encourage creativity whenever possible, or with respect to creativity, can there be too much of a good thing? Are the needs of the individual for creative expression at odds with society's need to reinforce conventions and established protocols? EVOC agents are too rudimentary to suffer the affective penalties of creativity but the model incorporates another drawback to creativity: time spent inventing is time not spent imitating. Creative agents, whose efforts go into reprocessing their own ideas rather than copying others, effectively rupture the fabric of the artificial society; they act as insulators that impede the diffusion of proven solutions. Imitators, in contrast, serve as a ``cultural memory" that ensures the preservation of successful ideas. When effective inventors created more and poor inventors created less, the society as a whole could capitalize on the creative abilities of the best inventors and capitalize on efforts of the rest to disseminate fit cultural outputs. The results suggest that it can be beneficial for a social group if individuals are allowed to follow different developmental trajectories in accordance with their demonstrated successes. 

These results do not prove that in real societies successful creators invent more and unsuccessful creators invent less; they merely show this kind of self-regulation is a feasible means of increasing the mean fitness of creative outputs. However, the fact that strong individual differences in creativity exist \cite{Kaufman2003,WolfradtPretz2001} suggests that this occurs in real societies. Whether prompted by individuals themselves or mediated by way of social cues, families, organizations, or societies may evolve faster by spontaneously self-organizing to achieve a balance between creative processes that generate innovations and the imitative processes that disseminate these innovations. They thereby temper novelty with continuity. 
%If this is the case, expensive and widely used programs to enhance creativity through methods such as brainstorming may be counterproductive. 
A more complex version of this scheme is that individuals find a task at which they excel, such that for each task domain there exists some individual in the social group who comes to be best equipped to explore that space of possibilities.

The social practice of discouraging creativity until the individual has proven him- or herself may serve as a form of social self-regulation ensuring that creative efforts are not squandered. 
Individuals who are tuned to social norms and expectations may over time become increasingly concerned with imitating and cooperating with others in a manner that promotes cultural continuity. Their thoughts travel more well-worn routes, and they are increasingly less likely to innovate. 
Others might be tuned to the demands of creative tasks, and less tethered to social norms and expectations, and thereby more likely to see things from unconventional perspectives. Thus they are more likely to come up with solutions to problems or unexpected challenges, find new avenues for self-expression, and contribute to the generation of cultural novelty.  In other words, what Cropley \emph{et al.} (2010) refer to as the ``dark side of creativity" may reflect that the creative individual is tuned to task needs at expense of human needs. Although in the long run this benefits the group as a whole because it results in creative outputs, in the short run the creative individual may be less likely to obey social norms and live up to social expectations, and to experience stigmatization or discrimination as a result, particularly in his/her early years \cite{Craft2005,Scott1999,Torrance1963}. Once the merits of such individuals' creative efforts become known, they may be supported or even idolized. 

The goal here was to investigate a hypothesis concerning the relationship between creativity and society rather than to come up with the most realistic model of creativity possible. However, a limitation of this work is that currently EVOC does not allow an agent to imitate some features of an idea and not others. This would be useful because cultural outputs both in EVOC and the real world exhibit a version of what in biology is referred to as epistasis, wherein what is optimal with respect to one component depends on what is going on with respect to another. Once both components have been optimized in a mutually beneficial way (in EVOC, for example, symmetrical arm movement), excess creativity risks breaking up co-adapted partial solutions. In future studies we will investigate the effects of enabling partial imitation. Another limitation is that the fitness function was static throughout a run, and agents had only one action to optimize. In real life, there are many tasks, and a division of labor such that each agent specializes in a few tasks, and imitates other agents to carry out other tasks. It may be that no one individual is an across-the-board ``creator" or ``imitator" but that different individuals find different niches for domain-specific creative outputs.

\section{Acknowledgments}

This work was supported by grants from the Natural Sciences and Engineering Research Council of Canada and the Flemish Fund for Scientific Research, Belgium.

%\vspace{-0.2cm}
% \small

% below commented out to prevent double Reference titles
%\section*{References}
%\begin{description}
 %\setlength{\itemsep}{-1mm}

\bibliographystyle{apacite}

\setlength{\bibleftmargin}{.125in}
\setlength{\bibindent}{-\bibleftmargin}

\bibliography{CogSci_Template}

\begin{thebibliography}{}

\bibitem [\protect \citeauthoryear {%
Andreason%
}{%
Andreason%
}{%
{\protect \APACyear {1987}}%
}]{%
Andreason1987}
\APACinsertmetastar {%
Andreason1987}%
\begin{APACrefauthors}%
Andreason, N.%
\end{APACrefauthors}%
\unskip\
\newblock
\APACrefYearMonthDay{1987}{}{}.
\newblock
{\BBOQ}\APACrefatitle {Creativity and mental illness. Prevalence rates in
  writers and their first degree relatives} {Creativity and mental illness.
  prevalence rates in writers and their first degree relatives}.{\BBCQ}
\newblock
\APACjournalVolNumPages{American Journal of Psychiatry}{144}{}{1288--1292}.
\PrintBackRefs{\CurrentBib}

\bibitem [\protect \citeauthoryear {%
Batey%
\ \BBA {} Furnham%
}{%
Batey%
\ \BBA {} Furnham%
}{%
{\protect \APACyear {2006}}%
}]{%
BateyFurnham2006}
\APACinsertmetastar {%
BateyFurnham2006}%
\begin{APACrefauthors}%
Batey, M.%
\BCBT {}\ \BBA {} Furnham, A.%
\end{APACrefauthors}%
\unskip\
\newblock
\APACrefYearMonthDay{2006}{}{}.
\newblock
{\BBOQ}\APACrefatitle {Creativity, intelligence, and personality: A critical
  review of the scattered literature} {Creativity, intelligence, and
  personality: A critical review of the scattered literature}.{\BBCQ}
\newblock
\APACjournalVolNumPages{Genetic and Social General Psychology
  Monographs}{7}{}{355--429}.
\PrintBackRefs{\CurrentBib}

\bibitem [\protect \citeauthoryear {%
Best%
}{%
Best%
}{%
{\protect \APACyear {1999}}%
}]{%
Best1999}
\APACinsertmetastar {%
Best1999}%
\begin{APACrefauthors}%
Best, M.%
\end{APACrefauthors}%
\unskip\
\newblock
\APACrefYearMonthDay{1999}{}{}.
\newblock
{\BBOQ}\APACrefatitle {How culture can guide evolution: An inquiry into
  gene/meme enhancement and opposition} {How culture can guide evolution: An
  inquiry into gene/meme enhancement and opposition}.{\BBCQ}
\newblock
\APACjournalVolNumPages{Adaptive Behavior}{132}{}{289--293}.
\PrintBackRefs{\CurrentBib}

\bibitem [\protect \citeauthoryear {%
Craft%
}{%
Craft%
}{%
{\protect \APACyear {2005}}%
}]{%
Craft2005}
\APACinsertmetastar {%
Craft2005}%
\begin{APACrefauthors}%
Craft, A.%
\end{APACrefauthors}%
\unskip\
\newblock
\APACrefYear{2005}.
\newblock
\APACrefbtitle {Creativity in schools: Tensions and dilemmas} {Creativity in
  schools: Tensions and dilemmas}.
\newblock
\APACaddressPublisher{London}{Routledge}.
\PrintBackRefs{\CurrentBib}

\bibitem [\protect \citeauthoryear {%
Cropley%
, Cropley%
, Kaufman%
\BCBL {}\ \BBA {} Runco%
}{%
Cropley%
\ \protect \BOthers {.}}{%
{\protect \APACyear {2010}}%
}]{%
CropleyCropleyKaufmanRunco2010}
\APACinsertmetastar {%
CropleyCropleyKaufmanRunco2010}%
\begin{APACrefauthors}%
Cropley, D.%
, Cropley, A.%
, Kaufman, J.%
\BCBL {}\ \BBA {} Runco, M.%
\end{APACrefauthors}%
\unskip\
\newblock
\APACrefYear{2010}.
\newblock
\APACrefbtitle {The dark side of creativity} {The dark side of creativity}.
\newblock
\APACaddressPublisher{Cambridge UK}{Cambridge University Press}.
\PrintBackRefs{\CurrentBib}

\bibitem [\protect \citeauthoryear {%
Eysenck%
}{%
Eysenck%
}{%
{\protect \APACyear {1993}}%
}]{%
Eysenck1993}
\APACinsertmetastar {%
Eysenck1993}%
\begin{APACrefauthors}%
Eysenck, H.%
\end{APACrefauthors}%
\unskip\
\newblock
\APACrefYearMonthDay{1993}{}{}.
\newblock
{\BBOQ}\APACrefatitle {Creativity and personality: Suggestions for a theory}
  {Creativity and personality: Suggestions for a theory}.{\BBCQ}
\newblock
\APACjournalVolNumPages{Psychological Inquiry}{4}{}{147--178}.
\PrintBackRefs{\CurrentBib}

\bibitem [\protect \citeauthoryear {%
Flaherty%
}{%
Flaherty%
}{%
{\protect \APACyear {2005}}%
}]{%
Flaherty2005}
\APACinsertmetastar {%
Flaherty2005}%
\begin{APACrefauthors}%
Flaherty, A.%
\end{APACrefauthors}%
\unskip\
\newblock
\APACrefYearMonthDay{2005}{}{}.
\newblock
{\BBOQ}\APACrefatitle {Frontotemporal and dopaminergic control of idea
  generation and creative drive} {Frontotemporal and dopaminergic control of
  idea generation and creative drive}.{\BBCQ}
\newblock
\APACjournalVolNumPages{Journal of Computational Neurology}{493}{}{147--153}.
\PrintBackRefs{\CurrentBib}

\bibitem [\protect \citeauthoryear {%
Florida%
}{%
Florida%
}{%
{\protect \APACyear {2002}}%
}]{%
Florida2002}
\APACinsertmetastar {%
Florida2002}%
\begin{APACrefauthors}%
Florida, R.%
\end{APACrefauthors}%
\unskip\
\newblock
\APACrefYear{2002}.
\newblock
\APACrefbtitle {The rise of the creative class} {The rise of the creative
  class}.
\newblock
\APACaddressPublisher{London}{Basic Books}.
\PrintBackRefs{\CurrentBib}

\bibitem [\protect \citeauthoryear {%
Gabora%
}{%
Gabora%
}{%
{\protect \APACyear {1995}}%
}]{%
Gabora1995}
\APACinsertmetastar {%
Gabora1995}%
\begin{APACrefauthors}%
Gabora, L.%
\end{APACrefauthors}%
\unskip\
\newblock
\APACrefYearMonthDay{1995}{}{}.
\newblock
{\BBOQ}\APACrefatitle {Meme and Variations: A computational model of cultural
  evolution} {Meme and variations: A computational model of cultural
  evolution}.{\BBCQ}
\newblock
\BIn{} L.~Nadel\ \BBA {} D.~Stein\ (\BEDS), \APACrefbtitle {1993 Lectures in
  Complex Systems.} {1993 lectures in complex systems.}
\newblock
\APACaddressPublisher{Reading MA}{Addison-Wesley}.
\PrintBackRefs{\CurrentBib}

\bibitem [\protect \citeauthoryear {%
Gabora%
, Chia%
\BCBL {}\ \BBA {} Firouzi%
}{%
Gabora%
\ \protect \BOthers {.}}{%
{\protect \APACyear {2013}}%
}]{%
GaboraChiaFirouzi2013}
\APACinsertmetastar {%
GaboraChiaFirouzi2013}%
\begin{APACrefauthors}%
Gabora, L.%
, Chia, W.%
\BCBL {}\ \BBA {} Firouzi, H.%
\end{APACrefauthors}%
\unskip\
\newblock
\APACrefYearMonthDay{2013}{}{}.
\newblock
{\BBOQ}\APACrefatitle {A computational model of two cognitive transitions
  underlying cultural evolution} {A computational model of two cognitive
  transitions underlying cultural evolution}.{\BBCQ}
\newblock
\BIn{} \APACrefbtitle {Proceedings of the 35th Annual Meeting of the Cognitive
  Science Society} {Proceedings of the 35th annual meeting of the cognitive
  science society}\ (\BPGS\ 2344--2349).
\newblock
\APACaddressPublisher{Houston TX}{Cognitive Science Society}.
\PrintBackRefs{\CurrentBib}

\bibitem [\protect \citeauthoryear {%
Gabora%
\ \BBA {} Firouzi%
}{%
Gabora%
\ \BBA {} Firouzi%
}{%
{\protect \APACyear {2012}}%
}]{%
GaboraFirouzi2012}
\APACinsertmetastar {%
GaboraFirouzi2012}%
\begin{APACrefauthors}%
Gabora, L.%
\BCBT {}\ \BBA {} Firouzi, H.%
\end{APACrefauthors}%
\unskip\
\newblock
\APACrefYearMonthDay{2012}{}{}.
\newblock
{\BBOQ}\APACrefatitle {Society functions best with an intermediate level of
  creativity} {Society functions best with an intermediate level of
  creativity}.{\BBCQ}
\newblock
\BIn{} \APACrefbtitle {Proceedings of the 34th Annual Meeting of the Cognitive
  Science Society} {Proceedings of the 34th annual meeting of the cognitive
  science society}\ (\BPGS\ 1578--1583).
\newblock
\APACaddressPublisher{Houston TX}{Cognitive Science Society}.
\PrintBackRefs{\CurrentBib}

\bibitem [\protect \citeauthoryear {%
Gino%
\ \BBA {} Ariely%
}{%
Gino%
\ \BBA {} Ariely%
}{%
{\protect \APACyear {2012}}%
}]{%
GinoAriely2012}
\APACinsertmetastar {%
GinoAriely2012}%
\begin{APACrefauthors}%
Gino, F.%
\BCBT {}\ \BBA {} Ariely, D.%
\end{APACrefauthors}%
\unskip\
\newblock
\APACrefYearMonthDay{2012}{}{}.
\newblock
{\BBOQ}\APACrefatitle {The dark side of creativity: Original thinkers can be
  more dishonest} {The dark side of creativity: Original thinkers can be more
  dishonest}.{\BBCQ}
\newblock
\APACjournalVolNumPages{Journal of Personality and Social
  Psychology}{102}{}{445--459}.
\PrintBackRefs{\CurrentBib}

\bibitem [\protect \citeauthoryear {%
Goodwin%
}{%
Goodwin%
}{%
{\protect \APACyear {1998}}%
}]{%
Goodwin1998}
\APACinsertmetastar {%
Goodwin1998}%
\begin{APACrefauthors}%
Goodwin, D.%
\end{APACrefauthors}%
\unskip\
\newblock
\APACrefYear{1998}.
\newblock
\APACrefbtitle {Alcohol and the Writer} {Alcohol and the writer}.
\newblock
\APACaddressPublisher{New York}{Penguin}.
\PrintBackRefs{\CurrentBib}

\bibitem [\protect \citeauthoryear {%
Guilford%
}{%
Guilford%
}{%
{\protect \APACyear {1950}}%
}]{%
Guilford1950}
\APACinsertmetastar {%
Guilford1950}%
\begin{APACrefauthors}%
Guilford, J.%
\end{APACrefauthors}%
\unskip\
\newblock
\APACrefYearMonthDay{1950}{}{}.
\newblock
{\BBOQ}\APACrefatitle {Creativity} {Creativity}.{\BBCQ}
\newblock
\APACjournalVolNumPages{American Psychologisty}{5}{}{444--454}.
\PrintBackRefs{\CurrentBib}

\bibitem [\protect \citeauthoryear {%
Higgs%
}{%
Higgs%
}{%
{\protect \APACyear {1992}}%
}]{%
Higgs2000}
\APACinsertmetastar {%
Higgs2000}%
\begin{APACrefauthors}%
Higgs, P.%
\end{APACrefauthors}%
\unskip\
\newblock
\APACrefYearMonthDay{1992}{}{}.
\newblock
{\BBOQ}\APACrefatitle {The mimetic transition: a simulation study of the
  evolution of learning by imitation} {The mimetic transition: a simulation
  study of the evolution of learning by imitation}.{\BBCQ}
\newblock
\APACjournalVolNumPages{Proceedings of the Royal Society B - Biological
  Sciences}{267}{}{1355--1361}.
\PrintBackRefs{\CurrentBib}

\bibitem [\protect \citeauthoryear {%
Hinton%
\ \BBA {} Nowlan%
}{%
Hinton%
\ \BBA {} Nowlan%
}{%
{\protect \APACyear {1992}}%
}]{%
HintonNowlan1987}
\APACinsertmetastar {%
HintonNowlan1987}%
\begin{APACrefauthors}%
Hinton, G.%
\BCBT {}\ \BBA {} Nowlan, S.%
\end{APACrefauthors}%
\unskip\
\newblock
\APACrefYearMonthDay{1992}{}{}.
\newblock
{\BBOQ}\APACrefatitle {How learning can guide evolution} {How learning can
  guide evolution}.{\BBCQ}
\newblock
\APACjournalVolNumPages{Complex Systems}{267}{}{495--502}.
\PrintBackRefs{\CurrentBib}

\bibitem [\protect \citeauthoryear {%
Holland%
}{%
Holland%
}{%
{\protect \APACyear {1975}}%
}]{%
Holland1975}
\APACinsertmetastar {%
Holland1975}%
\begin{APACrefauthors}%
Holland, J.%
\end{APACrefauthors}%
\unskip\
\newblock
\APACrefYear{1975}.
\newblock
\APACrefbtitle {Adaptation in Natural and Artificial Systems} {Adaptation in
  natural and artificial systems}.
\newblock
\APACaddressPublisher{Ann Arbor}{University of Michigan Press}.
\PrintBackRefs{\CurrentBib}

\bibitem [\protect \citeauthoryear {%
Hutchins%
\ \BBA {} Hazelhurst%
}{%
Hutchins%
\ \BBA {} Hazelhurst%
}{%
{\protect \APACyear {1991}}%
}]{%
HutchinsHazelhurst1991}
\APACinsertmetastar {%
HutchinsHazelhurst1991}%
\begin{APACrefauthors}%
Hutchins, E.%
\BCBT {}\ \BBA {} Hazelhurst, B.%
\end{APACrefauthors}%
\unskip\
\newblock
\APACrefYearMonthDay{1991}{}{}.
\newblock
{\BBOQ}\APACrefatitle {Learning in the cultural process} {Learning in the
  cultural process}.{\BBCQ}
\newblock
\BIn{} C.~Langton, J.~Taylor, D.~Farmer\BCBL {}\ \BBA {} S.~Rasmussen\ (\BEDS),
  \APACrefbtitle {Artificial life II.} {Artificial life ii.}
\newblock
\APACaddressPublisher{Redwood City}{Addison-Wesley}.
\PrintBackRefs{\CurrentBib}

\bibitem [\protect \citeauthoryear {%
Jamison%
}{%
Jamison%
}{%
{\protect \APACyear {1993}}%
}]{%
Jamieson1993}
\APACinsertmetastar {%
Jamieson1993}%
\begin{APACrefauthors}%
Jamison, K.%
\end{APACrefauthors}%
\unskip\
\newblock
\APACrefYear{1993}.
\newblock
\APACrefbtitle {Touched by fire: Manic-depressive illness and the artistic
  temperament} {Touched by fire: Manic-depressive illness and the artistic
  temperament}.
\newblock
\APACaddressPublisher{New York}{Free Press}.
\PrintBackRefs{\CurrentBib}

\bibitem [\protect \citeauthoryear {%
Kaufman%
}{%
Kaufman%
}{%
{\protect \APACyear {2003}}%
}]{%
Kaufman2003}
\APACinsertmetastar {%
Kaufman2003}%
\begin{APACrefauthors}%
Kaufman, J.%
\end{APACrefauthors}%
\unskip\
\newblock
\APACrefYearMonthDay{2003}{}{}.
\newblock
{\BBOQ}\APACrefatitle {The cost of the muse: Poets die young} {The cost of the
  muse: Poets die young}.{\BBCQ}
\newblock
\APACjournalVolNumPages{Death Studies}{27}{}{813--822}.
\PrintBackRefs{\CurrentBib}

\bibitem [\protect \citeauthoryear {%
Ludwig%
}{%
Ludwig%
}{%
{\protect \APACyear {1995}}%
}]{%
Ludwig1995}
\APACinsertmetastar {%
Ludwig1995}%
\begin{APACrefauthors}%
Ludwig, A.%
\end{APACrefauthors}%
\unskip\
\newblock
\APACrefYear{1995}.
\newblock
\APACrefbtitle {The Price of Greatness} {The price of greatness}.
\newblock
\APACaddressPublisher{New York}{Guilford Press}.
\PrintBackRefs{\CurrentBib}

\bibitem [\protect \citeauthoryear {%
Maslow%
}{%
Maslow%
}{%
{\protect \APACyear {1959}}%
}]{%
Maslow1959}
\APACinsertmetastar {%
Maslow1959}%
\begin{APACrefauthors}%
Maslow, A.%
\end{APACrefauthors}%
\unskip\
\newblock
\APACrefYearMonthDay{1959}{}{}.
\newblock
{\BBOQ}\APACrefatitle {Creativity in self-actualizing people} {Creativity in
  self-actualizing people}.{\BBCQ}
\newblock
\BIn{} Harper\ \BBA {} Brothers\ (\BEDS), \APACrefbtitle {Creativity and its
  cultivation.} {Creativity and its cultivation.}
\newblock
\APACaddressPublisher{New York}{McGraw-Hill}.
\PrintBackRefs{\CurrentBib}

\bibitem [\protect \citeauthoryear {%
May%
}{%
May%
}{%
{\protect \APACyear {1975}}%
}]{%
May1975}
\APACinsertmetastar {%
May1975}%
\begin{APACrefauthors}%
May, R.%
\end{APACrefauthors}%
\unskip\
\newblock
\APACrefYear{1975}.
\newblock
\APACrefbtitle {The courage to create} {The courage to create}.
\newblock
\APACaddressPublisher{New York}{Bantam}.
\PrintBackRefs{\CurrentBib}

\bibitem [\protect \citeauthoryear {%
Moran%
}{%
Moran%
}{%
{\protect \APACyear {2011}}%
}]{%
Moran2011}
\APACinsertmetastar {%
Moran2011}%
\begin{APACrefauthors}%
Moran, S.%
\end{APACrefauthors}%
\unskip\
\newblock
\APACrefYearMonthDay{2011}{}{}.
\newblock
{\BBOQ}\APACrefatitle {The roles of creativity in society} {The roles of
  creativity in society}.{\BBCQ}
\newblock
\BIn{} J.~Kaufman\ \BBA {} R.~Sternberg\ (\BEDS), \APACrefbtitle {Cambridge
  handbook of creativity.} {Cambridge handbook of creativity.}
\newblock
\APACaddressPublisher{Cambridge UK}{Cambridge University Press}.
\PrintBackRefs{\CurrentBib}

\bibitem [\protect \citeauthoryear {%
Qian%
, Plucker%
\BCBL {}\ \BBA {} Shen%
}{%
Qian%
\ \protect \BOthers {.}}{%
{\protect \APACyear {2010}}%
}]{%
QianPluckerShen2010}
\APACinsertmetastar {%
QianPluckerShen2010}%
\begin{APACrefauthors}%
Qian, M.%
, Plucker, J.%
\BCBL {}\ \BBA {} Shen, J.%
\end{APACrefauthors}%
\unskip\
\newblock
\APACrefYearMonthDay{2010}{}{}.
\newblock
{\BBOQ}\APACrefatitle {A model of Chinese adolescents� creative personality}
  {A model of chinese adolescents� creative personality}.{\BBCQ}
\newblock
\APACjournalVolNumPages{Creativity Research Journal}{22}{}{62--67}.
\PrintBackRefs{\CurrentBib}

\bibitem [\protect \citeauthoryear {%
Rogers%
}{%
Rogers%
}{%
{\protect \APACyear {1959}}%
}]{%
Rogers1959}
\APACinsertmetastar {%
Rogers1959}%
\begin{APACrefauthors}%
Rogers, C.%
\end{APACrefauthors}%
\unskip\
\newblock
\APACrefYearMonthDay{1959}{}{}.
\newblock
{\BBOQ}\APACrefatitle {Toward a theory of creativity} {Toward a theory of
  creativity}.{\BBCQ}
\newblock
\BIn{} H.~Anderson\ (\BED), \APACrefbtitle {Creativity and its cultivation.}
  {Creativity and its cultivation.}
\newblock
\APACaddressPublisher{New York}{Harper \& Row}.
\PrintBackRefs{\CurrentBib}

\bibitem [\protect \citeauthoryear {%
Rothenberg%
}{%
Rothenberg%
}{%
{\protect \APACyear {1990}}%
}]{%
Rothenberg1990}
\APACinsertmetastar {%
Rothenberg1990}%
\begin{APACrefauthors}%
Rothenberg, A.%
\end{APACrefauthors}%
\unskip\
\newblock
\APACrefYearMonthDay{1990}{}{}.
\newblock
{\BBOQ}\APACrefatitle {Creativity, mental health, and alcoholism} {Creativity,
  mental health, and alcoholism}.{\BBCQ}
\newblock
\APACjournalVolNumPages{Creativity Research Journal}{3}{}{179--201}.
\PrintBackRefs{\CurrentBib}

\bibitem [\protect \citeauthoryear {%
Scott%
}{%
Scott%
}{%
{\protect \APACyear {1999}}%
}]{%
Scott1999}
\APACinsertmetastar {%
Scott1999}%
\begin{APACrefauthors}%
Scott, C.%
\end{APACrefauthors}%
\unskip\
\newblock
\APACrefYearMonthDay{1999}{}{}.
\newblock
{\BBOQ}\APACrefatitle {Teachers� biases toward creative children}
  {Teachers� biases toward creative children}.{\BBCQ}
\newblock
\APACjournalVolNumPages{Creativity Research Journal}{12}{}{321--337}.
\PrintBackRefs{\CurrentBib}

\bibitem [\protect \citeauthoryear {%
Sternberg%
\ \BBA {} Lubart%
}{%
Sternberg%
\ \BBA {} Lubart%
}{%
{\protect \APACyear {1995}}%
}]{%
SternbergLubart1995}
\APACinsertmetastar {%
SternbergLubart1995}%
\begin{APACrefauthors}%
Sternberg, R.%
\BCBT {}\ \BBA {} Lubart, T.%
\end{APACrefauthors}%
\unskip\
\newblock
\APACrefYear{1995}.
\newblock
\APACrefbtitle {Defying the crowd: Cultivating creativity in a culture of
  conformity} {Defying the crowd: Cultivating creativity in a culture of
  conformity}.
\newblock
\APACaddressPublisher{New York}{Free Press}.
\PrintBackRefs{\CurrentBib}

\bibitem [\protect \citeauthoryear {%
Sulloway%
}{%
Sulloway%
}{%
{\protect \APACyear {1996}}%
}]{%
Sulloway1996}
\APACinsertmetastar {%
Sulloway1996}%
\begin{APACrefauthors}%
Sulloway, F.%
\end{APACrefauthors}%
\unskip\
\newblock
\APACrefYear{1996}.
\newblock
\APACrefbtitle {Born to rebel} {Born to rebel}.
\newblock
\APACaddressPublisher{New York}{Pantheon}.
\PrintBackRefs{\CurrentBib}

\bibitem [\protect \citeauthoryear {%
Tacher%
\ \BBA {} Readdick%
}{%
Tacher%
\ \BBA {} Readdick%
}{%
{\protect \APACyear {2006}}%
}]{%
TacherReaddick2006}
\APACinsertmetastar {%
TacherReaddick2006}%
\begin{APACrefauthors}%
Tacher, E.%
\BCBT {}\ \BBA {} Readdick, C.%
\end{APACrefauthors}%
\unskip\
\newblock
\APACrefYearMonthDay{2006}{}{}.
\newblock
{\BBOQ}\APACrefatitle {The relation between aggression and creativity among
  second graders} {The relation between aggression and creativity among second
  graders}.{\BBCQ}
\newblock
\APACjournalVolNumPages{Creativity Research Journal}{18}{}{261�267}.
\PrintBackRefs{\CurrentBib}

\bibitem [\protect \citeauthoryear {%
Tomasello%
, Kruger%
\BCBL {}\ \BBA {} Ratner%
}{%
Tomasello%
\ \protect \BOthers {.}}{%
{\protect \APACyear {1993}}%
}]{%
Tomasello1993}
\APACinsertmetastar {%
Tomasello1993}%
\begin{APACrefauthors}%
Tomasello, M.%
, Kruger, A.%
\BCBL {}\ \BBA {} Ratner, H.%
\end{APACrefauthors}%
\unskip\
\newblock
\APACrefYearMonthDay{1993}{}{}.
\newblock
{\BBOQ}\APACrefatitle {Cultural learning} {Cultural learning}.{\BBCQ}
\newblock
\APACjournalVolNumPages{Behavioral and Brain Sciences}{16}{}{495--552}.
\PrintBackRefs{\CurrentBib}

\bibitem [\protect \citeauthoryear {%
Torrance%
}{%
Torrance%
}{%
{\protect \APACyear {1963}}%
}]{%
Torrance1963}
\APACinsertmetastar {%
Torrance1963}%
\begin{APACrefauthors}%
Torrance, E.%
\end{APACrefauthors}%
\unskip\
\newblock
\APACrefYear{1963}.
\newblock
\APACrefbtitle {Guiding creative talent} {Guiding creative talent}.
\newblock
\APACaddressPublisher{Englewood Cliffs, NJ}{Prentice-Hall}.
\PrintBackRefs{\CurrentBib}

\bibitem [\protect \citeauthoryear {%
Treffinger%
, Young%
, Selby%
\BCBL {}\ \BBA {} Shepardson%
}{%
Treffinger%
\ \protect \BOthers {.}}{%
{\protect \APACyear {2002}}%
}]{%
TreffingerYoungShelbyShepardson2002}
\APACinsertmetastar {%
TreffingerYoungShelbyShepardson2002}%
\begin{APACrefauthors}%
Treffinger, D.%
, Young, G.%
, Selby, E.%
\BCBL {}\ \BBA {} Shepardson, C.%
\end{APACrefauthors}%
\unskip\
\newblock
\APACrefYear{2002}.
\newblock
\APACrefbtitle {Assessing creativity: A guide for educators (RM02170)}
  {Assessing creativity: A guide for educators (rm02170)}.
\newblock
\APACaddressPublisher{Storrs CT}{University of Connecticut Press and The
  National Research Center on the Gifted and Talented}.
\PrintBackRefs{\CurrentBib}

\bibitem [\protect \citeauthoryear {%
Troyer%
\ \BBA {} Youngreen%
}{%
Troyer%
\ \BBA {} Youngreen%
}{%
{\protect \APACyear {2009}}%
}]{%
TroyerYoungreen2009}
\APACinsertmetastar {%
TroyerYoungreen2009}%
\begin{APACrefauthors}%
Troyer, L.%
\BCBT {}\ \BBA {} Youngreen, R.%
\end{APACrefauthors}%
\unskip\
\newblock
\APACrefYearMonthDay{2009}{}{}.
\newblock
{\BBOQ}\APACrefatitle {Conflict and creativity in groups} {Conflict and
  creativity in groups}.{\BBCQ}
\newblock
\APACjournalVolNumPages{Journal of Social Issues}{65}{}{409--413}.
\PrintBackRefs{\CurrentBib}

\bibitem [\protect \citeauthoryear {%
Wolfradt%
\ \BBA {} Pretz%
}{%
Wolfradt%
\ \BBA {} Pretz%
}{%
{\protect \APACyear {2001}}%
}]{%
WolfradtPretz2001}
\APACinsertmetastar {%
WolfradtPretz2001}%
\begin{APACrefauthors}%
Wolfradt, U.%
\BCBT {}\ \BBA {} Pretz, J.%
\end{APACrefauthors}%
\unskip\
\newblock
\APACrefYearMonthDay{2001}{}{}.
\newblock
{\BBOQ}\APACrefatitle {Individual differences in creativity: Personality, story
  writing, and hobbies} {Individual differences in creativity: Personality,
  story writing, and hobbies}.{\BBCQ}
\newblock
\APACjournalVolNumPages{European Journal of Personality}{15}{}{297--310}.
\PrintBackRefs{\CurrentBib}

\end{thebibliography}

%\end{description}
\end{document}